# LATTICE FLUCTUATIONS SPECTRUM CALCULATION FOR P-DIHLORBENZOL NANOCRYSTALS

M.A.Korshunov

*L.V.Kirensky Institute of Physics Siberian Branch of the Russian Academy of Sciences, 660036 Krasnoyarsk, Russia.*

E-mail: kors@iph.krasn.ru

**Abstract** We have measured spectra of lattice fluctuations in p-dihlorbenzol nanoparticles of ~300nm size. Calculations of lattice frequencies and of these fluctuations in nanoparticles are done. It is shown that with reduction of nanoparticles sizes the spectrum of superficial fluctuations becomes prevailing.

Organic crystals find the increasing application in molecular electronics. Are thus used nanotechnologie. Transition from massive crystals to nanoparticles is accompanied by change of the periods of a lattice and orientation of molecules [1]. It should find reflexion in change of dynamics of a lattice and accordingly in spectra lattice fluctuations. The surface role increases at reduction of the sizes of crystals. For studying of this question comparison of experimental spectra of combinational dispersion of light lattice fluctuations nanoparticles p-dihlorbenzol with settlement spectra nanoparticles has been spent. For spectrum interpretation lattice fluctuations calculations on a method the Dyne [2] are carried out. In work are investigated nanoparticles p-dihlorbenzol. Massive single crystals p-dihlorbenzol are well enough studied by various methods and there is their interpretation lattice fluctuations [3].

On a glass plate has been raised dust in vacuum p-dihlorbenzol. The size nanoparticles was defined by an electronic microscope (~300nm). After that spectra of combinational dispersion of light of small frequencies have been received.

The spectrum lattice the fluctuations, received for nanoparticles p-dihlorbenzol, is resulted in figure 1 (a). The single crystal p-dihlorbenzol crystallises in spatial group $P2_1/a$ with two molecules in an elementary cell [4]. In a spectrum lattice single crystal fluctuations it should be observed six intensive lines caused rotary качаниями molecules round the main moments of inertia. The experimental spectrum is resulted in figure 1b. P - ($C_6H_4Cl_2$, ν, cm-1 single crystals 27.0, 47.0, 48.0, 55.0, 92.5, 101.0. p-$C_6H_4Cl_2$, ν, cm$^{-1}$ ~ 400 nanometers 22.0, 40.0, 44.0, 51.0, 84.0, 92.0, 12.0, 17.0, 23.0, 31.0, 63.0, 72.0, 110.0).

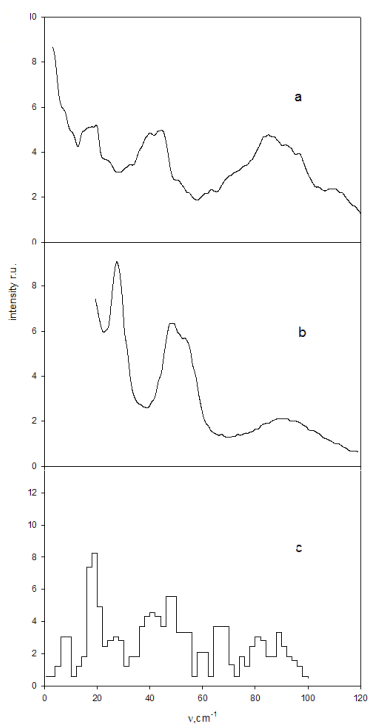

Figure 1. An experimental spectrum nanoparticles p-dihlorbenzol (~300nm) (a), a single crystal (b) and the calculated histogram of a spectrum.

In a spectrum it is observed six intensive lines and six lines of smaller intensity.

For an explanation of observable changes of frequencies of lines calculations of spectra lattice fluctuations on a method the Dyne [2] have been carried out. Factors in the potential of interaction were are taken from work [5]. On the basis of calculations histograms which show probability of display of lines of a spectrum in the chosen frequency interval have been received.

For nanoparticles p-dihlorbenzol minimisation on energy was spent at a various arrangement of molecules. Change of parametres of a lattice at change of the sizes nanoparticles has been found. Calculations have shown that the structure nanoparticles keeps the considered size a single crystal lattice p-dihlorbenzol though superficial molecules have ориентационную and transmitting disorder. With reduction of the sizes of particles lattice parametres increase.

In work results of calculations lattice and superficial fluctuations nanoparticles p-dihlorbenzol free of defects are resulted. Calculations have shown that the most intensive lines are connected with rotary качаниями molecules round the main moments of inertia. In figures 2 (a, b, c) histograms of spectra of frequencies are shown at change of the sizes nanoparticles.

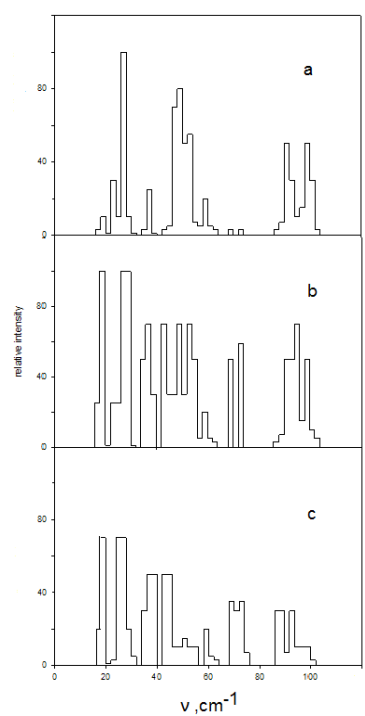

Figure 2. Histograms of spectra nanoparticles p-dihlorbenzol.

Calculations have shown that to reduction of the sizes of crystals there is an increase of a role of a spectrum of superficial fluctuations in spectrum nanoparticles. Calculation has been carried out at ориентационной disorder of superficial molecules. It has given occurrence of additional lines in area below 15 cm$^{-1}$ and other lines of a spectrum, but in the field of 80 cm$^{-1}$ of lines has not appeared (Figure 2). At the account of vacancies of a line in this area have appeared. In figure 1a the experimental spectrum nanoparticles p-dihlorbenzol (~300nm) and the calculated histogram of a spectrum at the account ориентационной is shown disorder of superficial molecules and presence of vacancies (fig. 1c). Apparently the settlement spectrum is close to an experimental spectrum.

Thus, it is visible that lines lattice fluctuations nanoparticles move to raising that, apparently, is caused by increase in the period of a lattice. With reduction of the sizes nanoparticles the spectrum of superficial fluctuations becomes prevailing. For correct interpretation of experimental spectra it is necessary to consider ориентационную disorder of superficial molecules and presence of vacancies in volume nanoparticles that, apparently, it is reflected in occurrence of some additional lines of smaller intensity.